\documentclass{PoS}

\usepackage{amssymb,amsmath,textcomp}
\usepackage{graphicx} 
\usepackage[caption=false]{subfig}	%for subfloats, option for clash with captions in documentclass
\usepackage{slashed}

\newcommand{\fermi}{\,\mathrm{fm}}
\newcommand{\brackets}[1]{\langle \,#1\,\rangle}
\newcommand{\xvec}{\vec{x}}
\newcommand{\yvec}{\vec{y}}
\newcommand{\Khat}{\hat{K}}

\title{Using analytic continuation for the
hadronic vacuum polarization computation}

\ShortTitle{Using analytic continuation for the
hadronic vacuum polarization computation}

\author{Xu Feng\\
       High Energy Accelerator Research Organization (KEK), Tsukuba 305-0801, Japan\footnote{present address: 
       Physics Department, Columbia University, New York, NY 10027, USA}\\
       E-mail: \email{pkufengxu@gmail.com}}

\author{Shoji Hashimoto\\
       High Energy Accelerator Research Organization (KEK), Tsukuba 305-0801, Japan\\
       E-mail: \email{shoji.hashimoto@kek.jp}}

\author{Grit Hotzel\\
       Humboldt University Berlin; NIC, Desy Zeuthen\\
       E-mail: \email{grit.hotzel@physik.hu-berlin.de}}

\author{\speaker{Karl Jansen}\\
       NIC, Desy Zeuthen, Platanenallee 6, 15738 Zeuthen, Germany\\
       Departament of Physics, University of Cyprus, P.O. Box 20537, 1678 Nicosia, Cyprus\\
       E-mail: \email{karl.jansen@desy.de}}

\author{Marcus Petschlies \\
       The Cyprus Institute, P.O. Box 27456, 1645 Nicosia, Cyprus \\
       E-mail: \email{m.petschlies@cyi.ac.cy}}

\author{Dru Renner \\
       Jefferson Lab, 12000 Jefferson Avenue, Newport News VA 23606, USA \\
       E-mail: \email{dru@jlab.org}}

\abstract{We present two examples of applications of the analytic continuation 
method for computing the hadronic vacuum polarization function in space- and time-like 
momentum regions. 
These examples are the Adler function and the leading order hadronic contribution to
the muon anomalous magnetic moment. We comment on the feasibility of the 
analytic continuation method and provide an outlook for possible further 
applications.} 

\FullConference{31st International Symposium on Lattice Field Theory LATTICE 2013\\
                 July 29 -- August 3, 2013\\
                 Mainz, Germany}

\begin{document}
\section{Introduction}

The calculation of the leading order hadronic contribution of the 
muon anomalous magnetic moment, $a_\mu^{\rm had}$, is one of the prime targets of 
lattice QCD activities presently. However, in such a computation 
there is a
generic problem to reach small momenta, dominating the weight function, 
on the lattice which are needed 
to evaluate the 
hadronic vacuum polarization (HVP) function from which 
$a_\mu^{\rm had}$ is derived. 
Present approaches to circumvent this problem  
\cite{Blum:2002ii,Gockeler:2003cw,Aubin:2006xv,Feng:2011zk,Boyle:2011hu,DellaMorte:2011aa,Aubin:2012me,deDivitiis:2012vs}
design appropriate fit functions for the HVP function, employ 
twisted boundary conditions, take 
model independent Pad\'e polynomials or compute 
the derivative of the vector current correlation function.

An alternative approach which we will discuss here is to use
the method of 
{\em analytic continuation} \cite{Feng:2013xsa} which is closely 
related to the work in refs.~\cite{Ji:2001wha,Meyer:2011um}. 
This method allows, in principle, to compute the HVP function at small 
space-like momenta and even at time-like momenta. We will show the feasibility of the method 
here at the examples of $a_\mu^{\rm had}$ and the Adler function. 
However, it is clear that the analytic continuation method is 
applicable for a much larger class of observables such as momentum dependent
form factors and quantities related to scattering processes or resonances. 

\section{Method of analytic continuation}

In the analytic continuation method the 
HVP function is computed through the Fourier transformation of the 
vector correlation function, 

\begin{equation}
\bar{\Pi}(K^2)(K_\mu K_\nu-\delta_{\mu\nu}K^2)= \int dt\;e^{\omega t}\int d^3\vec{x}\;e^{i\vec{k}\vec{x}}
\;\langle\Omega| T\{ J^{E}_\mu(\vec{x},t) J^{E}_\nu (\vec{0},0)\}|\Omega\rangle
\label{eq:FT}
\end{equation}
where 
${J^{E}_\mu(X)}$ is the electromagnetic current, 
${K=(\vec{k},-i\omega)}$ with $\vec{k}$ the spatial momentum and  
$\omega$ the photon energy.                          
The 
advantage of this approach is that 
$\omega$ can assume continuous
values for ${K^2=-\omega^2+\vec{k}^2}$ and thus allows to cover 
space-like and time-like momentum regions. In particular, it becomes 
possible to 
reach small momenta and even zero momentum. An 
important condition to be satisfied is 
$-K^2=\omega^2-\vec{k}^2<M_V^2$, 
with $M_V$ the invariant mass of the lowest
energy state in the vector channel, 
or $\omega<E_{\rm vector}$.  
In \cite{Ji:2001wha,Feng:2013xsa} a demonstration of the method has been given. 
In \cite{Feng:2013xsa} also the details of the calculation of the HVP function 
through the conserved vector correlation functions and a classification 
through the momenta $\vec{k}=(2\pi/L)\vec{n}$ has been provided. 

In practical calculations, a lattice with finite extent ${T}$ has to be used leading 
thus to possible finite size effects. In addition, for $t\gtrsim T/2$ the vector correlator 
becomes  
very noisy. In order to obtain a large signal to noise ratio, a cut in $t$ with a                                 
choice of ${t_{\rm max}=\eta(T/2)}$ has to be made for which we have fixed ${\eta=3/4}$. 
In the following, we will assume that for $t>t_{\rm max}$ the ground state 
dominates leading, with a continuous photon energy, to a suppression factor 
$exp[-(E_{\rm vector}-\omega)\cdot t_{\rm max}]$.

\section{Results for the HVP}

For our results we use gauge field ensembles with 
$N_f=2$ and $N_f=2+1+1$ flavours of maximally 
twisted mass sea fermions \cite{Baron:2009wt,Baron:2010bv}
employing two values of the lattice spacing, various volumes and 
a number of quark masses reaching pion masses as low as 210 MeV. 
We refer to ref.~\cite{Feng:2013xsa} for further details 
of the ensembles used. 

\subsection{Vacuum polarization and Adler functions} 

As a first example of the application of the analytic continuation method 
we will discuss here the evaluation of the Adler function.
The definition of the vacuum polarization function 
via analytic continuation presented in the previous
section from lattice data can be used directly to extract the
additively renormalized vacuum polarization
function. When extrapolated to the physical point, 
this allows for an immediate comparison with dispersion relation results available
via experimental data for the ratio $R(s)$, involving the $e^{+}e^{-}$ cross section to hadrons. 
Furthermore, 
with the knowledge of 
$R(s)$ it is also possible to compute the Adler function 
which can then be compared to a lattice calculation.  
To this end, we use 
\begin{align}
\bar{\Pi}(K^2) &= \frac{1}{12}\,\sum\limits_{s}^{t_\mathrm{max}/a}\sum\limits_{\xvec}\,
                  \brackets{J_i(t_x,\xvec)\,J_i(t_y,\yvec)}\,P_s(\Khat^2/\Lambda^2)
        \label{eq:tensor_def_V}
\end{align}
with (Chebyshev based) polynomials $P_s$ that can be derived from the Fourier transformation 
of eq.~(\ref{eq:FT}). For any finite $T$ we thus have a representation of the polarization 
function as a sum over polynomials, which are summed up to some maximal order 
$t_\mathrm{max}$ as explained above. This reproduces the
lattice data on the points in momentum space, but the polynomials can also 
be evaluated for any choice of $\Khat^2$.
In particular, we can take the limit $\Khat^2 \to 0$ and choose $\Khat^2 = -\hat{\Omega}^2 < 0$ to smoothly connect
the space-like and (small) time-like momentum region.

\begin{figure}[htpb]
\begin{center}
\includegraphics[width=0.47\textwidth]{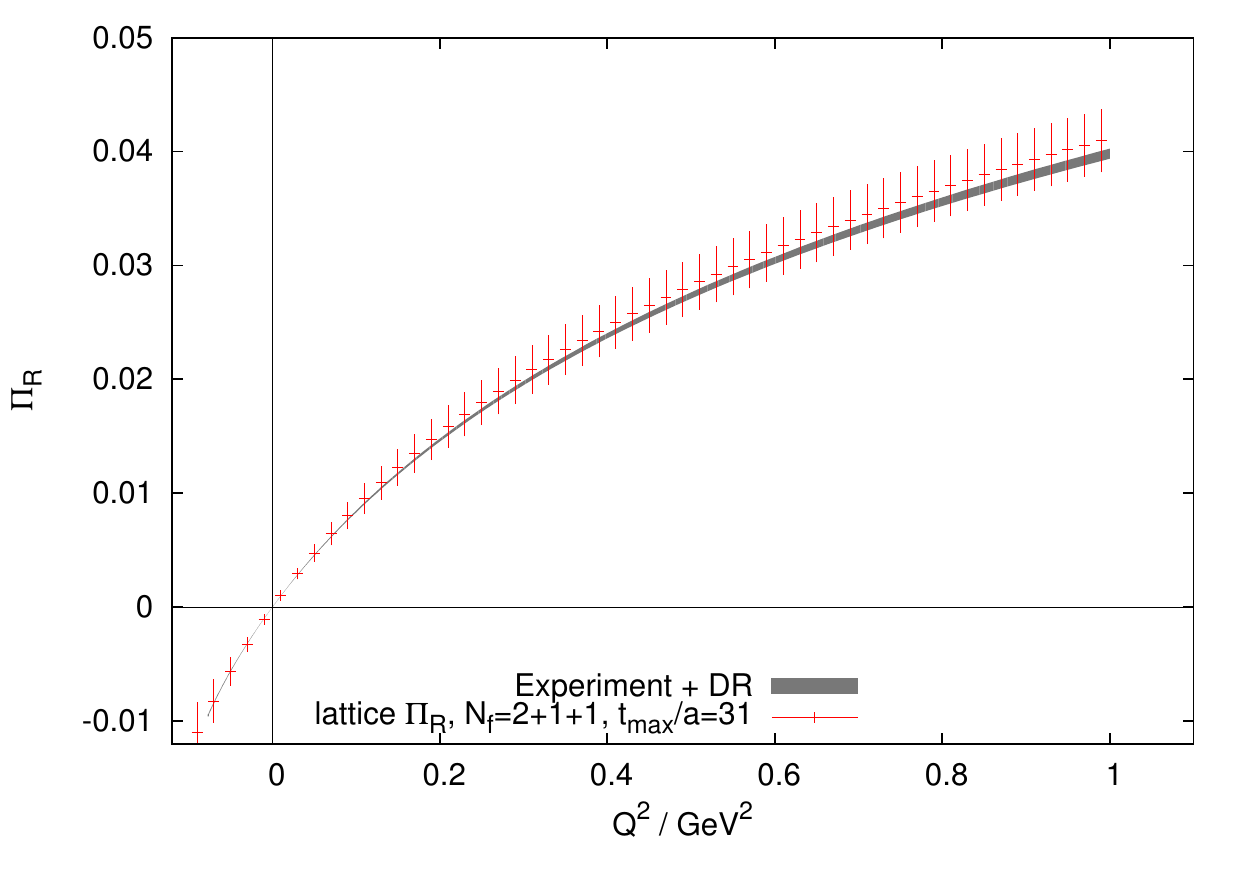}
\includegraphics[width=0.47\textwidth]{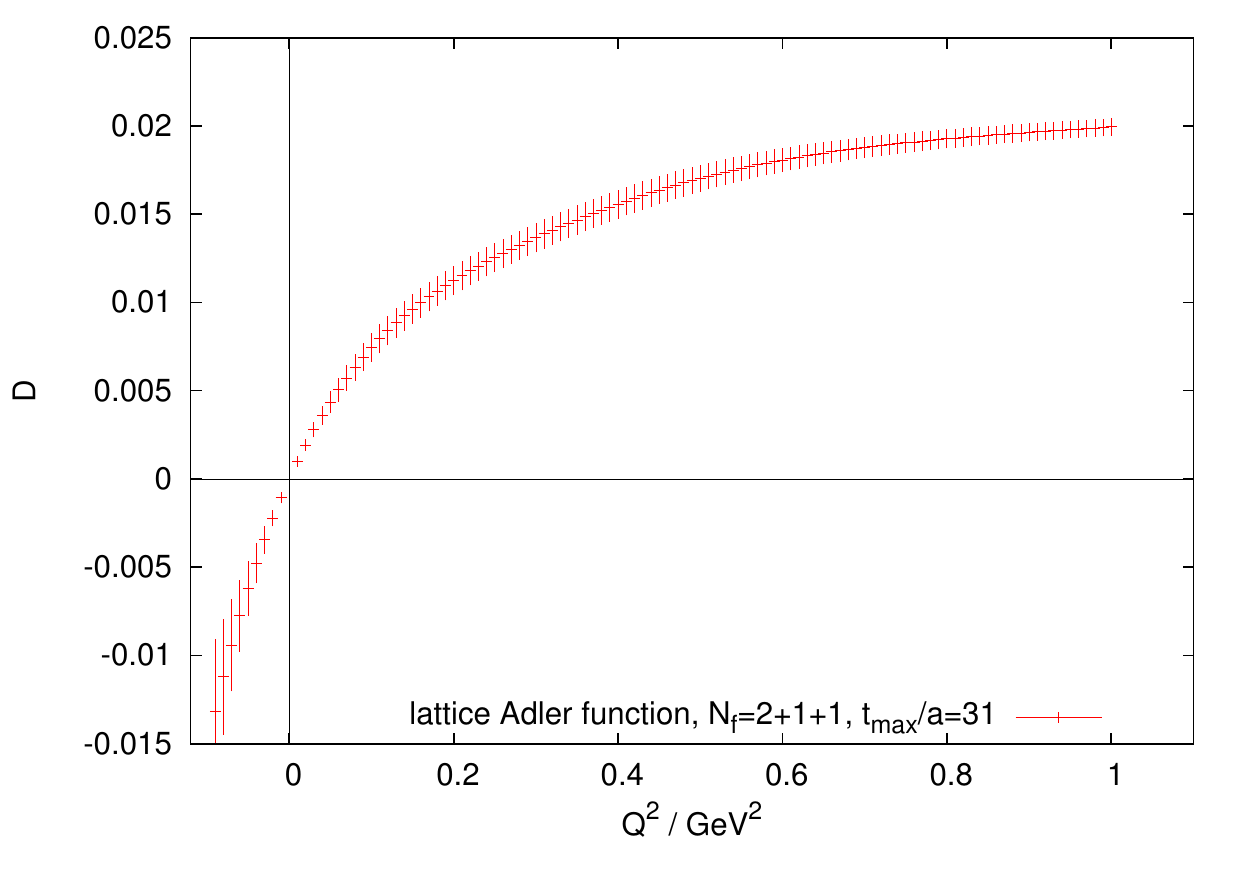}
\end{center}
\caption{left panel: The renormalized vacuum polarization function from our lattice 
        data ($a \approx 0.078\,\mathrm{fm},\,V = (2.5\fermi)^3$) together 
        with a comparison to the dispersion relation data from \cite{Jegerlehner:2011mw}. 
        right panel: The Adler function as derived from the lattice data; for illustration the upper summation limit
        was chosen maximal, i.e. $t_\mathrm{max}/a = T/2a - 1 = 31$ and the spatial momentum is set to 0.
}
        \label{fig:dr_latt_comparison}
\end{figure}

Here we compare our results, which
are obtained at a single lattice spacing ($a \approx 0.078\fermi$) with the results from \cite{Jegerlehner:2011mw} provided
in the package \textit{alphaQED}. At the stage of this first analysis, 
which is shown in the left panel of fig.~\ref{fig:dr_latt_comparison} we find compelling agreement of results
using the dispersion relation for $R(s)$ and lattice QCD results. With the known functional dependence on $\Khat^2$ of
the $P_n$ of eq.~(\ref{eq:tensor_def_V}) we can take the derivative w.r.t. $\Khat^2$ analytically and can also 
derive the Adler function from the lattice data.
The result based on the same modified extrapolation of ref.~\cite{Feng:2011zk} is 
shown in the right panel of Figure \ref{fig:dr_latt_comparison}.
Details of the limit $t_\mathrm{max}/a \to \infty $ as well as generalizations to different choices of
non-zero momentum components will be discussed in a forthcoming publication.

\subsection{Anomalous magnetic moment of the muon} 

Having obtained the HVP function, besides the Adler function a number of physical observables can be 
computed, see, e.g., Ref.\cite{Renner:2012fa}.
Here we will focus on 
the leading-order HVP correction to the muon anomalous magnetic moment,
$a_\mu^{\rm hvp}$, as an
example to study the practical feasibility of the proposed analytic
continuation method.

In lattice QCD $a_\mu^{\rm hvp}$ can be calculated through
\begin{equation}
a_\mu^{\rm hvp}=\alpha^2\int_0^{\infty} dK^2\; \frac{1}{K^2}f\left(\frac{K^2}{m_\mu^2}\right)(\Pi(K^2)-\Pi(0))\;,
\end{equation}
where $\alpha$ is the fine structure constant,
$m_\mu$ is the muon mass, and $f(K^2/m_\mu^2)$ is a known function~\cite{Blum:2002ii}.
To control the chiral extrapolation, we use a modified definition of $a_\mu^{\rm hvp}$  
proposed in Refs.~\cite{Feng:2011zk,Renner:2012fa},
\begin{equation}
\label{eq:amu}
a_{\bar{\mu}}^{\rm hvp}=\alpha^2\int_0^{\infty} dK^2\; \frac{1}{K^2}
f\left(\frac{K^2}{m_\mu^2}\frac{H_{\rm phys}^2}{H^2}\right)(\Pi(K^2)-\Pi(0))\;,
\end{equation}
with $H=M_V$, i.e. the {\em measured} vector meson mass on the lattice. 
When the pion mass approaches its physical value, this vector meson mass 
becomes the physical $\rho$-meson mass, $M_V=M_\rho$. Then also
$H=H_{\rm phys}$ and the modified definition, $a_{\bar{\mu}}^{\rm hvp}$
reproduces the value of $a_\mu^{\rm hvp}$ at the physical pion mass.

In the analytic continuation method, we can calculate the HVP function for a
continuous momentum region
$0<K^2<K_{\rm max}^2$, with $K_{\rm max}^2=\sum_{i=x,y,z}\hat{K}_i^2$
being the squared spatial momentum.
In practice
we split
 Eq.~(\ref{eq:amu}) into three parts:
\begin{eqnarray}
\label{eq:amu_split}
a_{\bar{\mu}}^{\rm hvp}&=&a_{\bar{\mu}}^{(1)}+a_{\bar{\mu}}^{(2)}+a_{\bar{\mu}}^{(3)}\;,\nonumber\\
a_{\bar{\mu}}^{(1)}&=&\alpha^2\int_0^{K_{\rm max}^2} dK^2\; \frac{1}{K^2}
f\left(\frac{K^2}{m_\mu^2}\frac{H_{\rm phys}^2}{H^2}\right)(\Pi(K^2)-\Pi(0))\;,\nonumber\\
a_{\bar{\mu}}^{(2)}&=&\alpha^2\int_{K_{\rm max}^2}^{\infty} dK^2\; \frac{1}{K^2}
f\left(\frac{K^2}{m_\mu^2}\frac{H_{\rm phys}^2}{H^2}\right)(\Pi(K_{\rm max}^2)-\Pi(0))\;,\nonumber\\
a_{\bar{\mu}}^{(3)}&=&\alpha^2\int_{K_{\rm max}^2}^{\infty} dK^2\; \frac{1}{K^2}
f\left(\frac{K^2}{m_\mu^2}\frac{H_{\rm phys}^2}{H^2}\right)(\Pi(K^2)-\Pi(K_{\rm max}^2))\;.
\end{eqnarray}
In Eq.~(\ref{eq:amu_split}), $a_{\bar{\mu}}^{(1)}+a_{\bar{\mu}}^{(2)}$
can be calculated directly using the analytic continuation method.
Similar to the previous section,
we can calculate $a_{\bar{\mu}}^{(1)}$ and $a_{\bar{\mu}}^{(2)}$ up to
the value of $t_{\rm max}=\eta(T/2)$ with $\eta=3/4$ and estimate the
finite-size effects using $\bar{\Pi}(K^2;t>t_{\rm max})$.

The evaluation of $a_{\bar{\mu}}^{(3)}$ still requires a parametrization of $\bar{\Pi}(K^2)$.
This will bring in some model dependence in our analysis, which, however,
is a small effect, since the total contribution of $a_{\bar{\mu}}^{(3)}$ only
amounts to a few percent in case of momentum modes $|\vec{n}|^2=1,2,3$ where 
$\vec{n}$ classifies the momenta $\vec{k}=(2\pi/L)\vec{n}$ used in eq.~(\ref{eq:FT}).
The exact parametrization of $\bar{\Pi}(K^2)$ used in this calculation can 
be found in the Appendix of ref.~\cite{Feng:2013xsa}.

\begin{figure}[htpb]
\begin{center}
\includegraphics[width=300pt,angle=-90]{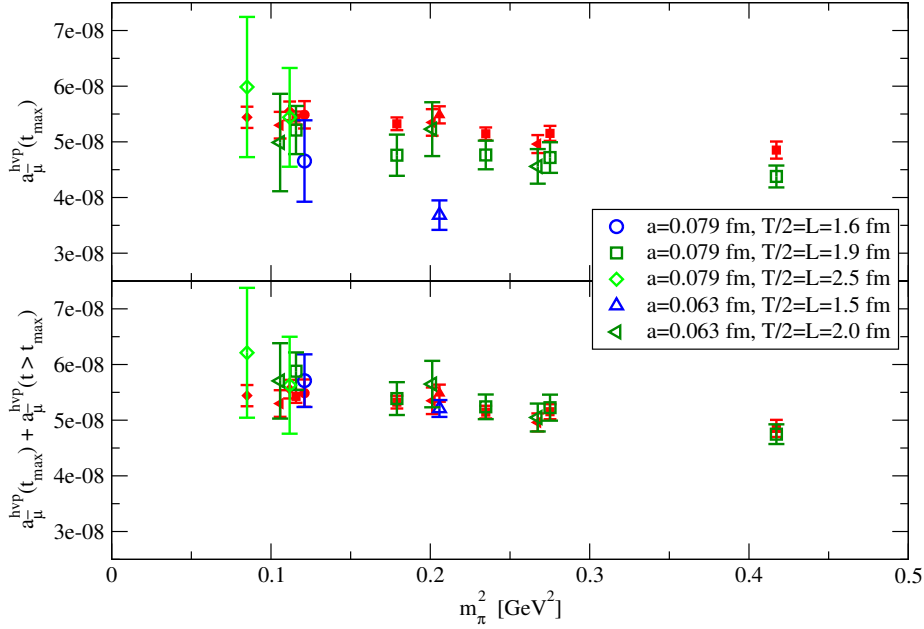}
\end{center}
\caption{ 
In this figure, the open symbols represent results from the analytic continuation
method where we average 
over {$|n|^2=1,2,3$}. In the          
upper panel we use the range $-t_{\rm max} \le t \le t_{\rm max}$ only. 
The
lower panel shows the results corrected for the estimated FSE.
Our earlier data from the standard method of 
computing $a_{\mu}^{\rm hvp}$ are represented by filled symbols. 
}
\label{fig:etacut}
\end{figure}

In Fig.~\ref{fig:etacut} we show $a_{\bar{\mu}}^{\rm hvp}$ as a function of the squared pion mass.
The results of $a_{\bar{\mu}}^{\rm hvp}$ calculated using Eq.~(\ref{eq:amu_split}) are shown
by the empty symbols. These results have been averaged among various momentum modes ($|\vec{n}|^2=1,2,3$) and polarization directions.
For comparison the results determined using our conventional approach of
parameterizing the HVP function in the full momentum range are shown in
the same figure, represented by the filled symbols.
In the upper panel we show the results of
$a_{\bar{\mu}}^{\rm hvp}(t_{\rm max})$, which are calculated using the correlator $C_{\mu\nu}(\vec{k},t)$ covering the range of
$-t_{\rm max} \le t\le t_{\rm max}$. We look at the finite-size effects
in $a_{\bar{\mu}}^{\rm hvp}(t_{\rm max})$
by comparing the results for different volumes. There are also some deviations between the
results from the analytic continuation method and the standard parametrization method.
To check for the finite-size effects, we evaluate the contribution to $a_{\bar{\mu}}^{\rm hvp}$
from $C_{\mu\nu}(\vec{k},t)$ at $|t|>t_{\rm max}$ leading to a correction
$a_{\bar{\mu}}^{\rm hvp}(t>t_{\rm max})$. The corresponding results for
$a_{\bar{\mu}}^{\rm hvp}(t_{\rm max})+a_{\bar{\mu}}^{\rm hvp}(t>t_{\rm max})$
are shown in the lower panel of Fig.~\ref{fig:etacut}. These results are consistent now
among different lattice
volumes. Besides this, the results from the analytic continuation method also agree with the ones from
the standard parametrization method.
As can be seen,
the results from the
analytic continuation method for $a_{\bar{\mu}}^{\rm hvp}$
show larger fluctuations than the standard ones. However, the analytic
continuation method has the conceptual advantage that
in the region of low $K^2$ the parametrization of
$\bar{\Pi}(K^2)$ can be avoided.
Thus, we think that presently the analytic
continuation method can serve as a valuable cross-check of the standard
method to analyze the vacuum polarization function.

\section{Conclusions and Outlook}

In this proceedings contribution we discussed the 
analytic continuation method which, in principle, can provide 
information on the small momentum region for the vacuum polarization function
or form factors in lattice QCD. In addition, it 
allows to access the space- and time-like momentum regions which 
in turn opens the possibility to compute scattering processes and 
resonances as an alternative to the finite size 
method of refs.~\cite{Luscher:1985dn,Luscher:1986pf,Luscher:1991cf}.
Here we have given the examples of the Adler function and the 
anomalous magnetic moment of the muon 
$a_{\mu}^{\rm hvp}$ which we calculated by the analytic continuation method. 
When comparing to the standard method to compute $a_{\mu}^{\rm hvp}$ we found full 
agreement, but the analytic continuation method leads to noisier results. 
Still, we believe that the analytic continuation method is a valuable alternative 
which has, moreover, the  
potential to address other quantities
where small or zero momenta are needed.

\begin{acknowledgments}
X.F. and S.H. are supported
in part by the Grant-in-Aid of the Japanese Ministry of Education (Grant No.\ 21674002), and
D.R. is supported in part by Jefferson Science Associates, LLC, under
U.S. DOE Contract No.\ DE-AC05-06OR23177.
This work has been supported in part by the DFG
Corroborative Research Center SFB/TR9.
G.H.~gratefully acknowledges the
support of the German Academic National Foundation (Studienstiftung des
deutschen Volkes e.V.) and of the DFG-funded Graduate School GK 1504.
G.H. gratefully acknowledges the support of the German Academic National Foundation
(Studienstiftung des deutschen Volkes e.V.).
K.J. is supported in part by the Cyprus Research Promotion
Foundation under Contract No. $\Pi$PO$\Sigma$E$\Lambda$KY$\Sigma$H/EM$\Pi$EIPO$\Sigma$/0311/16.
The numerical computations have been performed on the
{\it SGI system HLRN-II} at the {HLRN Supercomputing Service Berlin-Hannover},  FZJ/GCS,
and BG/P at FZ-J\"ulich.
\end{acknowledgments}

\bibliographystyle{unsrt}
% \addcontentsline{toc}{Chapter}{Bibliography}
% \clearpage
%\bibliography{lattice2013_references.bib}
\bibliography{Lattice_2013_proceedings.bib}
% \begin{thebibliography}{99}
%   \bibitem{...} ....
% \end{thebibliography}

\end{document}